\documentclass{elsart}

\usepackage{epsfig}

\usepackage{amssymb}

\begin{document}

\begin{frontmatter}



\title{First observation of generation in the backward wave oscillator with a "grid" diffraction grating   
and lasing of the volume FEL  
with a "grid" volume resonator}


\author{V.G. Baryshevsky},
\author{K.G.Batrakov,}
\author{N.A.Belous,}
\author{A.A.Gurinovich,}
\author{A.S.Lobko,}
\author{P.V.Molchanov,}
\author{P.F.Sofronov,}
\author{V.I.Stolyarsky.}

\address{Research Institute of Nuclear Problem, 
Belarussian State University,   \\   11 Bobruyskaya Str. , Minsk 220050, Belarus}

\begin{abstract}
First observation of generation in the backward wave oscillator with a "grid" diffraction grating   
and lasing of the volume FEL  
with a "grid" volume resonator
that is formed by a perodic set of metallic threads inside a rectangular waveguide is considered.
\end{abstract}


\end{frontmatter}

\qquad 
Generation of radiation in millimeter and far-infrared
range with nonrelativistic and low-relativistic electron beams
gives rise difficulties. Gyrotrons and cyclotron resonance
facilities are used as sources in millimeter and sub-millimeter
range, but for their operation magnetic field about several tens
of kiloGauss ($\omega \sim \frac{eH}{mc}\gamma $) is necessary.
Slow-wave devices (TWT, BWT, orotrons)in this range require
application of dense and thin ($<0.1$ mm) electron beams, because
only electrons passing near the slowing structure at the distance
$\leq \lambda \beta \gamma /(4\pi )$ can interact with
electromagnetic wave effectively.
It is difficult to guide thin beams near slowing structure with
desired accuracy. And electrical endurance of resonator limits
radiation power and density of acceptable electron beam.
Conventional waveguide systems are essentially restricted by the
requirement for transverse dimensions of resonator, which should
not significantly exceed radiation wavelength. Otherwise,
generation efficiency decreases abruptly due to excitation of
plenty of modes. The most of the above problems can be overpassed
in VFEL
\cite{PhysLett,VFELreview,FirstLasing,FEL2002,patent}. 
In VFEL the greater part of electron beam interacts with
the electromagnetic wave due to volume distributed interaction.
Transverse dimensions of VFEL resonator could significantly exceed
radiation wavelength $D \gg \lambda $. In addition, electron beam
and radiation power are distributed over the whole volume that is
beneficial for electrical endurance of the system.

The electrodynamical properties of volume
diffraction structures composed from strained dielectric threads was experimentally
studied in 
\cite{VolumeGrating}. In \cite{THz} it was shown that nonrelativistic and
low-relativistic electron beams passing through such structures
can generate in wide frequency range up to terahertz.
The electrodynamical properties of a "grid" volume resonator
formed by a perodic structure built
from the metallic threads inside a rectangular waveguide 
was considered in \cite{resonator}.

\begin{figure}[h]
\epsfxsize = 8 cm 
\centerline{\epsfbox{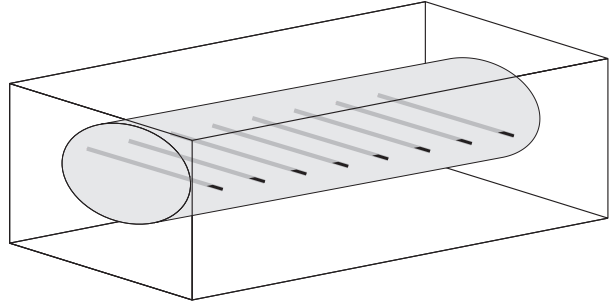}}
\caption{A backward wave oscillator with
the "grid" diffraction grating}
\label{one_grating}
\end{figure}

\qquad
In the present paper first observation of lasing of the backward wave oscillator with 
a "grid" diffraction grating   
and the volume FEL  
with a "grid" volume resonator
that is formed by a perodic set of metallic threads inside a rectangular waveguide 
(see Fig.\ref{one_grating})
is considered. 
According to the analysis \cite{resonator} a "grid" volume resonator was
built from tungsten threads with diameter 0.1 mm
strained inside the rectangular resonator with transverse dimensions 45 mm x 50 mm and length 300 mm 
(see Fig.\ref{two_gratings}). 
Distance between threads was 12.5 mm. 
Annular electron beam with the energy 200 keV and electron beam current 2kA passed through the above resonator.
As it was mentioned above only electrons passing near the slowing structure at the distance
$\delta \leq \lambda \beta \gamma /(4\pi )$ can interact with
electromagnetic wave effectively. In our case annular electron beam with the 
outer radius 16 mm and
the ring width 4 mm 
passes through
a "grid"
diffraction grating. 
As a result, the part of the electron beam moving close to the gratinge inside the interval 
$2\delta\}$ interacts with the grating effectively Fig.3.
The purpose of experiment was to prove possibility of lasing in 
such a resonator and to examine ''grid'' volume resonator durability.

Two different experiments were carried out (Fig.\ref{one_grating},Fig.\ref{two_gratings}):
the experiment providing to observe generation in backward wave oscillator with
a "grid" diffraction grating and lasing of the volume FEL  
with a "grid" volume resonator.

The first experiment was done to test the system as a backward wave oscillator.
Radiation pulse with power about 1 kWatt was registered.

Then additional diffraction grating (Fig.\ref{two_gratings}) was placed inside the resonator
to provide volume distributed feedback in the system.
Lasing of volume free electron laser with ''grid'' volume resonator was studied.
Radiation pulse with frequency ~10 GHz and power about 10 kWatt was registered.

\begin{figure}[h]
\epsfxsize = 8 cm 
\centerline{\epsfbox{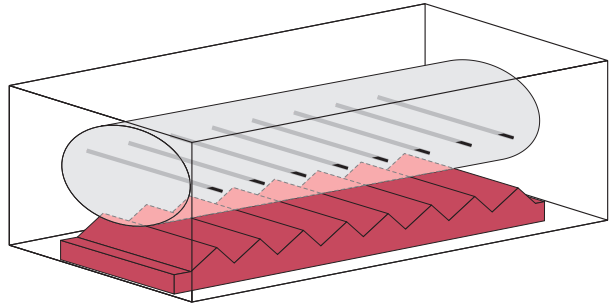}}
\caption{"Grid" volume resonator}
\label{two_gratings}
\end{figure}

\begin{figure}[h]
\epsfxsize = 8 cm 
\centerline{\epsfbox{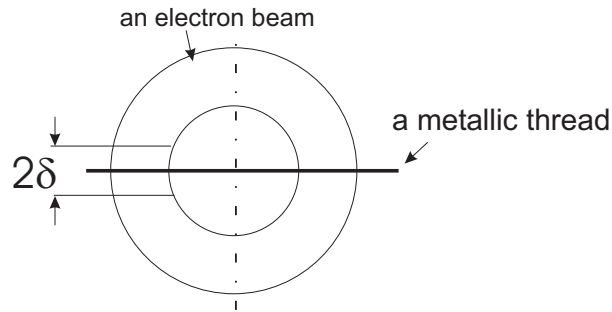}}
\caption{Annular electron beam interacts with "grid" grating}
\label{two_gratings}
\end{figure}

Thus, durability of ''grid'' volume resonator and possibility of lasing in 
such a resonator was proved.

The experiments with several parallel "grid" gratings are in progress now.


\end{document}